\documentclass[a4paper,fleqn,usenatbib]{mnras}

\usepackage{newtxtext,newtxmath}
\usepackage{subfigure}
\usepackage{threeparttable}
\usepackage{multirow}
\usepackage{remreset}
\usepackage{CJKutf8}

\usepackage[T1]{fontenc}
\usepackage{ae,aecompl}

\usepackage{graphicx}   
\usepackage{amsmath}    
\usepackage{amssymb}    
\usepackage{makecell}   
\usepackage{threeparttable} 
\usepackage{multirow}   

\title[Velocity shift of Mg II and Al III BALs]{Velocity shift of Mg II and Al III broad absorption lines in quasar SDSS J134444.33+315007.6}

\author[Lu\& Lin]{
Wei-Jian Lu\footnotemark[1] and
Ying-Ru Lin\footnotemark[2] 
\\
School of Information Engineering, Baise University, Baise 533000, China}

\date{Accepted XXX. Received YYY; in original form ZZZ}

\pubyear{2019}

\begin{document}
\label{firstpage}
\pagerange{\pageref{firstpage}--\pageref{lastpage}}
\maketitle

\begin{abstract}
We report, for the first time, a synchronized velocity shift of \ion{Mg}{II} and \ion{Al}{III} broad absorption lines (BALs) in quasar SDSS J134444.33+315007.6 (hereafter J1344+3150). {We found this quasar from a sample of 134 \ion{Mg}{II} BAL quasars with multi-epoch observations.} This quasar contains three low-ionization BAL systems, the fastest of which  at $\thicksim$$-17, 000\,\rm km\,s^{-1}$ shows a kinematic shift of $\thicksim$$-1101\,\rm km\,s^{-1}$ and $\thicksim$$-1170\,\rm km\,s^{-1}$ in its \ion{Mg}{II} and \ion{Al}{III} ions, respectively, during a rest-frame time of about 3.21 yr. Meanwhile, this quasar also shows other various variation characteristics, including an obvious weakening in its continuum, a coordinated enhancement in multiple emission lines (\ion{Mg}{II}, \ion{C}{III} and \ion{Al}{III}), and a coordinated enhancement in three \ion{Al}{III} absorption troughs. These variation characteristics convincingly indicate that the BAL outflows of J1344+3150 are under the influence from the background radiation energy. Thus, we infer that the velocity shift displayed in system A in the quasar J1344+3150 may indicate an actual line-of-sight acceleration of an outflow due to the radiation pressure from the central source.

\end{abstract}
%
\begin{keywords}
galaxies : active -- quasars : absorption lines -- quasars: individual (SDSS J1344+3150).
\end{keywords}

\footnotetext[1]{E-mail: william\_lo@qq.com (W-J L)}
\footnotetext[2]{E-mail: yingru\_lin@qq.com (Y-R L)}



\section{Introduction}
At present, {there is a lack of} understandings for the velocity shift of absorption lines seen in the ultraviolet (UV) spectra of quasars. This is in sharp contrast with the extensive research on the equivalent width (EW) variability of absorption lines. But we still cannot ignore that an observed velocity shift of an absorption profile can be a signature of potential acceleration/deceleration, which is an important physical property of the outflowing gas in quasars. 

On the one hand, the study of velocity shift of absorption lines is limited by the number of its direct detection cases. Recent studies have presented abundant cases or large samples to show the universality of the EW variations of absorption lines in quasars (for broad absorption lines, namely BALs: e.g. \citealp{Hamann2008,Grier2015,McGraw2017,He2017,He2019,Yi2019}, and references therein; for narrow absorption lines, namely NALs: e.g. \citealp{Hamann2011,Chen2015a,Chen2018a,Chen2018b}, and references therein). However, only a few studies have shown direct detection of the velocity shift of absorption lines (\citealp{Vilkoviskij2001,Rupke2002,Gabel2003,Hall2007,Joshi2014,Grier2016,Joshi2019,Lu2019shift,
Yao2020,Xu2020}). Most of these studies are based on single quasar objects, with the exception of \citet{Grier2016}, who firstly performed the systematic research of velocity shift using a quasar sample. In 140 BAL quasars that were observed at least three epochs by the Sloan Digital Sky Survey (SDSS; \citealp{York2000}), \citet{Grier2016} found only three quasars existing velocity shifts in the \ion{C}{IV} BALs. This means that, to date, there is no further quantitative statistical investigation or common trend analysis of the velocity shift of BALs that based on a large sample. Moreover, these existing studies of the velocity shift of absorption lines mainly focus on high-ionization the \ion{C}{IV} ion, sometimes along with the \ion{Si}{IV} ion, but there is no case on the velocity shift signatures in low-ionization ions, such as the \ion{Mg}{II} etc. 

\begin{figure*}
\centering
\includegraphics[width=2\columnwidth]{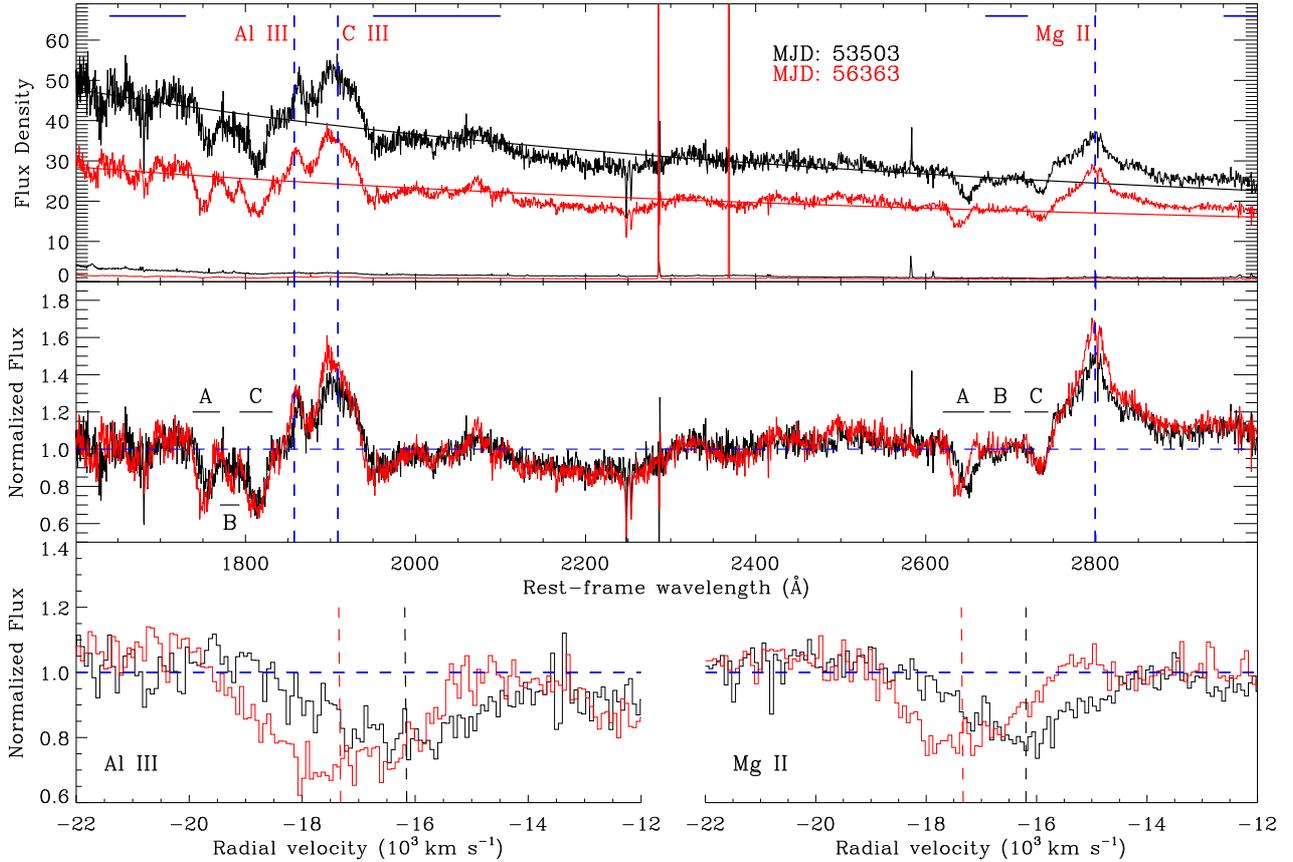}
\caption{Top panel: original spectra of quasar SDSS J1344+3150 that observed on MJD 53,503 (black) and 56,363 (red), together with their power-law continua. The blue vertical dashed lines represent the rest wavelengths of the main emission lines. {The y-axis is in unit of $10^{-17}$ erg cm$^{-2}$ s$^{-1}$.} The blue horizontal bars at the top are the wavelength regions we used to fit the power-law continua. Middle panel: normalized spectra of quasar J1344+3150 showing three BAL systems (marked by black horizontal solid bars); Bottom panel: snippets from normalized spectra showing the \ion{Mg}{II} and \ion{Al}{III} BALs of system A. The black (MJD 53,503) and red (MJD 56,363) horizontal vertical dashed lines are the line centers of the absorption troughs.}
\label{fig.1}
\end{figure*}

On the other hand, the research progress on the physical mechanism of the change in velocity of absorption lines has not kept up with EW variation. Though the cause of absorption trough EW variability is still debated in the literature, it is widely suggested that its possible causes include variations in the ionization state (e.g. \citealp{Hamann2011,He2017,Lu2017,LLQ2018,He2019}) and the transverse motion of the absorption materials across our line of sight (e.g. \citealp{Hamann2008}). Both these two origins have been supported by observational evidences. However, the cause of an observed velocity shift of an absorption line is still inconclusive, maybe a actual acceleration/deceleration of the outflows, a change in the line-of-sight velocity produced by directional shift of the outflows, or variations in velocity-dependent quantities, including in ionization state or covering factor (e.g. \citealp{Gabel2003,Hall2007,Grier2016,Joshi2019}). {Note that these causes for velocity shift of absorption lines have rare conclusive evidence. Recently, \citet{Xu2020} have presented a relatively reliable case that the BAL velocity shift is due to acceleration of an outflow, because they can exclude other alternate possibilities, including photoionization changes scenario and motion of outflow clouds into and out of the line of sight.} 

In this letter, we report, for the first time, a synchronized velocity shift of the \ion{Mg}{II} and \ion{Al}{III} BALs in quasar SDSS J134444.33+315007.6 ($z_{\rm em}=1.440$, hereafter J1344+3150). We found this special case based on a visually check of a sample of 134 \ion{Mg}{II} BAL quasars with multi-epoch observations by the SDSS, which was constructed by \citet{Yi2019}. 

The rest of this letter is divided into the following two parts. In Section \ref{sec.2} we present the methods and results of the spectra analysis and the measurements of broad absorption troughs. In Section \ref{sec.3} we discuss the implications of our results.

\begin{table*}
    \centering
\caption{Measurements of BALs.  \label{tab.1}}
\begin{tabular}{cccccccc} 
\hline 
\hline 
BAL Specie&  Velocity$^{\rm a}$ & FWHM$^{\rm b}$ & \multicolumn2c{EW(\AA)}  & Fractional  EW &Note\\
\cline{4-5}
 &($\rm km~s^{-1}$) &($\rm km~s^{-1}$) &MJD: 53,503& MJD: 56,363& Variation & \\
\hline
\ion{Mg}{II}	&	$	-19501 	\sim	-14009 	$	&	5493 	&	$	3.95 	\pm	0.25 	$	&	$	3.81 	\pm	0.25 	$	&	$	-0.04 	\pm	0.091 	$	&	system A	\\
\ion{Mg}{II}$\rm ^c$	&	$	-13292 	\sim	-10508 	$	&	2784 	&	$	0.60 	\pm	0.19 	$	&		——				&		——				&	system B	\\
\ion{Mg}{II}	&	$	-8738 	\sim	-5665 	$	&	3073 	&	$	1.69 	\pm	0.17 	$	&	$	1.35 	\pm	0.19 	$	&	$	-0.22 	\pm	0.172 	$	&	system C	\\
\ion{Al}{III}	&	$	-19493 	\sim	-14006 	$	&	5487 	&	$	3.65 	\pm	0.30 	$	&	$	5.17 	\pm	0.23 	$	&	$	0.35 	\pm	0.092 	$	&	system A	\\
\ion{Al}{III}	&	$	-14006 	\sim	-10142 	$	&	3864 	&	$	1.91 	\pm	0.26 	$	&	$	2.65 	\pm	0.20 	$	&	$	0.32 	\pm	0.151 	$	&	system B	\\
\ion{Al}{III}	&	$	-10142 	\sim	-3694 	$	&	6447 	&	$	7.08 	\pm	0.31 	$	&	$	8.66 	\pm	0.24 	$	&	$	0.20 	\pm	0.050 	$	&	system C	\\
\hline 
\end{tabular}
\begin{tablenotes}
\footnotesize
\item$^{\rm a}$Velocity range of the BAL trough that are calculated with respect to wavelength of the blue component of the absorption doublet.
\item$^{\rm b}$Total width calculated from edge-to-edge of the BAL trough.
\item$^{\rm c}$Note that \ion{Mg}{II} BAL in system B in the MJD 56,363 spectrum is nearly undetectable.
\end{tablenotes}
\end{table*}
\section{Spectra analysis and absorption trough detection} \label{sec.2} 
We obtained the spectra of quasar J1344+3150 from the SDSS data release 16 (DR16; \citealp{Ahumada2020}), which were archived on MJD 53,503 and MJD 56,363, respectively. The resolution of them is $R\approx2000$. The MJD 53,503 spectrum covers a wavelength range from 3800 to 9200\,\AA, while the MJD 56,363 spectrum covers a wavelength range from 3600 to 10,000\,\AA. The median signal-to-noise ratio (S/N) of the MJD 53,503 and MJD 56,363 spectra are {21 and 24}, respectively.

By iteratively fitting the continuum within the wavelength regions that are relatively free from strong emission lines and absorption lines, we obtained a power-law continuum of each original spectra (see the top panel of Figure \ref{fig.1}). Then we normalized the two original spectra using their corresponding power-law continuum (see the middle panel of Figure \ref{fig.1}). In the power-law continuum normalized spectra, we detected three broad absorption systems (systems A, B, and C). We measured the EW and full width at half-maximum (FWHM) of these three broad absorption systems in \ion{Al}{III} and \ion{Mg}{II} ions, following the methods in our previous studies (e.g. equations (2) and (3) in \citealp{Lu2018complex1}). The measurements can be seen in Table \ref{tab.1}. Note that system B in \ion{Mg}{II} ion in the MJD 56,363 spectrum is nearly undetectable. 

We found that system A exhibited a velocity shift between the two observations in both the \ion{Mg}{II} and \ion{Al}{III} ions. Following \citet{Lu2019shift}, we determined the line center of a BAL by averaging the EWs of the troughs. In order to more correctly estimate the line centers, we generated 10,000 simulated spectra via adding Gaussian noise to the original spectra according to the corresponding flux density errors, and then used the same procedure to measure the line centers for these simulated spectra. Then, we estimated the line center and the measurement errors according to the median values and the standard deviations of the 10,000 trials, respectively. Finally, we got the centroid radial velocities of the \ion{Mg}{II} BAL of system A for the MJD 53,503 and 56,363 spectra, which are $-16194\pm39\,\rm km\,s^{-1}$ and $-17295\pm61\,\rm km\,s^{-1}$, respectively. The difference between their velocities is $-1101\pm80\,\rm km\,s^{-1}$, which corresponds to an acceleration rate of $1.087\pm0.079\,\rm cm\,s^{-2}$. Meanwhile, the centroid radial velocities of the \ion{Al}{III} BAL of system A are $-16184\pm77\,\rm km\,s^{-1}$ and $-17353\pm29\,\rm km\,s^{-1}$ for the MJD 53,503 and 56,363 spectra, respectively, and their velocity difference is $-1170\pm69\,\rm km\,s^{-1}$, corresponding to an acceleration rate of $1.155\pm0.068\,\rm cm\,s^{-2}$.   

We integrated the wavelength range from 1600 to 3000\,\AA~to obtain the flux of each continuum, and find out that, during a rest-frame time of about 3.21 yr, the continuum of quasar J1344+3150 experienced an obvious weakening with the fraction variation of about $-0.423$.

\section{Discussion and conclusion} \label{sec.3}
As can be seen from Figure \ref{fig.1}, the BAL systems mainly exhibits two observational characteristics. One characteristic is the coordinated enhancement of all three \ion{Al}{III} absorption troughs. Besides, although suffering from large uncertainties, all three \ion{Mg}{II} troughs also show faint trends of weakening (see also Table \ref{tab.1}). Another characteristic is the synchronized velocity shift displayed in the fastest \ion{Mg}{II} and \ion{Al}{III} BALs (system A). 

First, we hold the view that the ionization change is more reasonable (than the transverse-motion scenario) to be the cause of the absorption line variation in J1344+3150, according to the following several reasons. For the first reason, variations in the continuum within our available wavelength coverage possibly indicate the changes in ionizing radiation at bluer wavelengths. For the second reason, significant variability of the broad emission line (BEL) generated in the photoionized broad-line region can be an indicator of a change in the ionizing radiation field, which may result in an ionization change of a BAL outflow. Last but not least, coordinated BAL variability covering a large velocity range ($\sim15,000\,\rm km\,s^{-1}$ in this source) can be a strong evidence in support of the ionization-change scenario, because this situation is difficult to be interpreted by the transverse-motion scenario, which requires multiple outflowing absorbers to move synchronously (see e.g. \citealp{Hamann2011,Grier2015,McGraw2017}).

Second, we suggest that the observed velocity shift may be related to the radiation pressure from the central source. As mentioned in the introduction, at present, the physical mechanism of the velocity shift of absorption line is still difficult to determine, and there is even no observational case of velocity shift that can be perfectly explained by existing models. For J1344+3150 in this letter, only relying on the data of two observations from the SDSS, we have not been able to determine the physical mechanism of its line velocity shift. However, between the two observations of this source, we detected the coordinated BAL variations cover a large velocity range, and significant fluctuations in the continuum and BELs. These variation characteristics convincingly indicate that the BAL outflow of J1344+3150 is under the influence from the background radiation energy. Thus, we infer that the velocity shift displayed in system A in quasar J1344+3150 may indicate an actual line-of-sight acceleration of an outflow due to radiation pressure from the central source. {If the actual acceleration of an outflow is indeed the reason for the BAL velocity shift displayed in J1344+3150, then its BAL measurements may be comparable with those of the J1042+1646 in \citet{Xu2020}, which BAL velocity shift is attributed to the same reason. First, both these two sources show a synchronized velocity shift of more than one ions, but the ion ionization states of J1042+1646 (\ion{Ne}{VIII}, \ion{O}{V} and \ion{Mg}{X}) are higher than J1344+3150 (\ion{Mg}{II} and \ion{Al}{III}); second, the outflow velocity of J1042+1646 ($\sim$$-20,000\,\rm km\,s^{-1}$) is faster than J1344+3150 ($\sim$$-17,000\,\rm km\,s^{-1}$); third, the velocity shift ($\sim$$-1550\,\rm km\,s^{-1}$) and average acceleration rate ($1.52\,\rm cm\,s^{-2}$) of J1042+1646 are both larger than those of J1344+3150. These comparisons may infer that more dramatic acceleration can be seen at higher ionization states and/or higher velocities.}

In conclusion, we have reported a velocity shift of a low-ionization broad absorption system (including \ion{Mg}{II} and \ion{Al}{III} ions) in quasar SDSS J1344+3150. According to the variation characteristics displayed in its two-epoch SDSS spectra, including an obvious weakening in its continuum, a coordinated enhancement in multiple emission lines (\ion{Mg}{II}, \ion{C}{III} and \ion{Al}{III}), and a coordinated enhancement in three \ion{Al}{III} absorption troughs, we infer that the velocity shift of the absorption lines shown in quasar J1344+3150 may be dominated by radiation pressure.

\section*{Acknowledgements}
We thank the reviewer for helpful comments. This research was supported by the National Natural Science Foundation of China (No. 11903002), the Guangxi Natural Science Foundation (No. 2017GXNSFAA198348), and the Research Project of Baise University (No. 2019KN04).

Funding for the Sloan Digital Sky Survey IV was
provided by the Alfred P. Sloan Foundation, the U.S.
Department of Energy Office of Science, and the Participating
Institutions. SDSS-IV acknowledges support and resources
from the Center for High-Performance Computing at the
University of Utah. The SDSS website is \url{http://www.sdss.org/}.

SDSS-IV is managed by the Astrophysical Research
Consortium for the Participating Institutions of the SDSS
Collaboration including the Brazilian Participation Group, the
Carnegie Institution for Science, Carnegie Mellon University,
the Chilean Participation Group, the French Participation
Group, Harvard-Smithsonian Center for Astrophysics, Instituto
de Astrofísica de Canarias, The Johns Hopkins University,
Kavli Institute for the Physics and Mathematics of the Universe
(IPMU)/University of Tokyo, Lawrence Berkeley National
Laboratory, Leibniz Institut für Astrophysik Potsdam (AIP),
Max-Planck-Institut für Astronomie (MPIA Heidelberg),
Max-Planck-Institut für Astrophysik (MPA Garching), MaxPlanck-Institut für Extraterrestrische Physik (MPE), National
Astronomical Observatories of China, New Mexico State
University, New York University, University of Notre Dame,
Observatário Nacional/MCTI, The Ohio State University,
Pennsylvania State University, Shanghai Astronomical Observatory, United Kingdom Participation Group, Universidad
Nacional Autónoma de México, University of Arizona,
University of Colorado Boulder, University of Oxford,
University of Portsmouth, University of Utah, University of
Virginia, University of Washington, University of Wisconsin,
Vanderbilt University, and Yale University.

\section*{DATA AVAILABILITY}
The data underlying this article were accessed from the Sloan Digital Sky Survey.

\bibliographystyle{mnras}
\bibliography{NALvsBALandshift} 

\begin{thebibliography}{}
\makeatletter
\relax
\def\mn@urlcharsother{\let\do\@makeother \do\$\do\&\do\#\do\^\do\_\do\%\do\~}
\def\mn@doi{\begingroup\mn@urlcharsother \@ifnextchar [ {\mn@doi@}
  {\mn@doi@[]}}
\def\mn@doi@[#1]#2{\def\@tempa{#1}\ifx\@tempa\@empty \href
  {http://dx.doi.org/#2} {doi:#2}\else \href {http://dx.doi.org/#2} {#1}\fi
  \endgroup}
\def\mn@eprint#1#2{\mn@eprint@#1:#2::\@nil}
\def\mn@eprint@arXiv#1{\href {http://arxiv.org/abs/#1} {{\tt arXiv:#1}}}
\def\mn@eprint@dblp#1{\href {http://dblp.uni-trier.de/rec/bibtex/#1.xml}
  {dblp:#1}}
\def\mn@eprint@#1:#2:#3:#4\@nil{\def\@tempa {#1}\def\@tempb {#2}\def\@tempc
  {#3}\ifx \@tempc \@empty \let \@tempc \@tempb \let \@tempb \@tempa \fi \ifx
  \@tempb \@empty \def\@tempb {arXiv}\fi \@ifundefined
  {mn@eprint@\@tempb}{\@tempb:\@tempc}{\expandafter \expandafter \csname
  mn@eprint@\@tempb\endcsname \expandafter{\@tempc}}}

\bibitem[\protect\citeauthoryear{{Ahumada} et~al.,}{{Ahumada}
  et~al.}{2020}]{Ahumada2020}
{Ahumada} R.,  et~al., 2020, \mn@doi [\apjs] {10.3847/1538-4365/ab929e}, \href
  {https://ui.adsabs.harvard.edu/abs/2020ApJS..249....3A} {249, 3}

\bibitem[\protect\citeauthoryear{{Chen}, {Gu}, {Chen}  \& {Cao}}{{Chen}
  et~al.}{2015}]{Chen2015a}
{Chen} Z.-F.,  {Gu} Q.-S.,  {Chen} Y.-M.,   {Cao} Y.,  2015, \mn@doi [\mnras]
  {10.1093/mnras/stv813}, \href
  {http://adsabs.harvard.edu/abs/2015MNRAS.450.3904C} {450, 3904}

\bibitem[\protect\citeauthoryear{{Chen}, {Pang}, {He}  \& {Huang}}{{Chen}
  et~al.}{2018a}]{Chen2018a}
{Chen} Z.-F.,  {Pang} T.-T.,  {He} B.,   {Huang} Y.,  2018a, \mn@doi [\apjs]
  {10.3847/1538-4365/aabcd4}, \href
  {http://adsabs.harvard.edu/abs/2018ApJS..236...39C} {236, 39}

\bibitem[\protect\citeauthoryear{{Chen} et~al.,}{{Chen}
  et~al.}{2018b}]{Chen2018b}
{Chen} Z.-F.,  et~al., 2018b, \mn@doi [\apjs] {10.3847/1538-4365/aaeac3}, \href
  {http://adsabs.harvard.edu/abs/2018ApJS..239...23C} {239, 23}

\bibitem[\protect\citeauthoryear{{Gabel} et~al.,}{{Gabel}
  et~al.}{2003}]{Gabel2003}
{Gabel} J.~R.,  et~al., 2003, \mn@doi [\apj] {10.1086/345096}, \href
  {http://cdsads.u-strasbg.fr/abs/2003ApJ...583..178G} {583, 178}

\bibitem[\protect\citeauthoryear{{Grier} et~al.,}{{Grier}
  et~al.}{2015}]{Grier2015}
{Grier} C.~J.,  et~al., 2015, \mn@doi [\apj] {10.1088/0004-637X/806/1/111},
  \href {http://cdsads.u-strasbg.fr/abs/2015ApJ...806..111G} {806, 111}

\bibitem[\protect\citeauthoryear{{Grier} et~al.,}{{Grier}
  et~al.}{2016}]{Grier2016}
{Grier} C.~J.,  et~al., 2016, \mn@doi [\apj] {10.3847/0004-637X/824/2/130},
  \href {http://adsabs.harvard.edu/abs/2016ApJ...824..130G} {824, 130}

\bibitem[\protect\citeauthoryear{{Hall}, {Sadavoy}, {Hutsemekers}, {Everett}
  \& {Rafiee}}{{Hall} et~al.}{2007}]{Hall2007}
{Hall} P.~B.,  {Sadavoy} S.~I.,  {Hutsemekers} D.,  {Everett} J.~E.,   {Rafiee}
  A.,  2007, \mn@doi [\apj] {10.1086/519273}, \href
  {http://adsabs.harvard.edu/abs/2007ApJ...665..174H} {665, 174}

\bibitem[\protect\citeauthoryear{{Hamann}, {Kaplan}, {Rodr{\'{\i}}guez
  Hidalgo}, {Prochaska}  \& {Herbert-Fort}}{{Hamann} et~al.}{2008}]{Hamann2008}
{Hamann} F.,  {Kaplan} K.~F.,  {Rodr{\'{\i}}guez Hidalgo} P.,  {Prochaska}
  J.~X.,   {Herbert-Fort} S.,  2008, \mn@doi [\mnras]
  {10.1111/j.1745-3933.2008.00554.x}, \href
  {http://cdsads.u-strasbg.fr/abs/2008MNRAS.391L..39H} {391, L39}

\bibitem[\protect\citeauthoryear{{Hamann}, {Kanekar}, {Prochaska}, {Murphy},
  {Ellison}, {Malec}, {Milutinovic}  \& {Ubachs}}{{Hamann}
  et~al.}{2011}]{Hamann2011}
{Hamann} F.,  {Kanekar} N.,  {Prochaska} J.~X.,  {Murphy} M.~T.,  {Ellison} S.,
   {Malec} A.~L.,  {Milutinovic} N.,   {Ubachs} W.,  2011, \mn@doi [\mnras]
  {10.1111/j.1365-2966.2010.17575.x}, \href
  {http://adsabs.harvard.edu/abs/2011MNRAS.410.1957H} {410, 1957}

\bibitem[\protect\citeauthoryear{{He}, {Wang}, {Zhou}, {Bian}, {Liu}, {Yang},
  {Dou}  \& {Sun}}{{He} et~al.}{2017}]{He2017}
{He} Z.,  {Wang} T.,  {Zhou} H.,  {Bian} W.,  {Liu} G.,  {Yang} C.,  {Dou} L.,
   {Sun} L.,  2017, \mn@doi [\apjs] {10.3847/1538-4365/aa647a}, \href
  {http://cdsads.u-strasbg.fr/abs/2017ApJS..229...22H} {229, 22}

\bibitem[\protect\citeauthoryear{{He} et~al.,}{{He} et~al.}{2019}]{He2019}
{He} Z.,  et~al., 2019, \mn@doi [Nature Astronomy] {10.1038/s41550-018-0669-8},
  \href {https://ui.adsabs.harvard.edu/abs/2019NatAs...3..265H} {3, 265}

\bibitem[\protect\citeauthoryear{{Joshi}, {Chand}, {Srianand}  \&
  {Majumdar}}{{Joshi} et~al.}{2014}]{Joshi2014}
{Joshi} R.,  {Chand} H.,  {Srianand} R.,   {Majumdar} J.,  2014, \mn@doi
  [\mnras] {10.1093/mnras/stu840}, \href
  {http://adsabs.harvard.edu/abs/2014MNRAS.442..862J} {442, 862}

\bibitem[\protect\citeauthoryear{{Joshi}, {Srianand}, {Chand}, {Wu},
  {Noterdaeme}, {Petitjean}  \& {Ho}}{{Joshi} et~al.}{2019}]{Joshi2019}
{Joshi} R.,  {Srianand} R.,  {Chand} H.,  {Wu} X.-B.,  {Noterdaeme} P.,
  {Petitjean} P.,   {Ho} L.~C.,  2019, \mn@doi [\apj]
  {10.3847/1538-4357/aaf500}, \href
  {http://adsabs.harvard.edu/abs/2019ApJ...871...43J} {871, 43}

\bibitem[\protect\citeauthoryear{{Lu} \& {Lin}}{{Lu} \&
  {Lin}}{2018}]{Lu2018complex1}
{Lu} W.-J.,  {Lin} Y.-R.,  2018, \mn@doi [\mnras] {10.1093/mnras/stx2970},
  \href {http://cdsads.u-strasbg.fr/abs/2018MNRAS.474.3397L} {474, 3397}

\bibitem[\protect\citeauthoryear{{Lu} \& {Lin}}{{Lu} \&
  {Lin}}{2019}]{Lu2019shift}
{Lu} W.-J.,  {Lin} Y.-R.,  2019, \mn@doi [\apj] {10.3847/1538-4357/ab53d7},
  \href {https://ui.adsabs.harvard.edu/abs/2019ApJ...887..178L} {887, 178}

\bibitem[\protect\citeauthoryear{{Lu} et~al.,}{{Lu} et~al.}{2017}]{Lu2017}
{Lu} W.-J.,  et~al., 2017, \mn@doi [\mnras] {10.1093/mnrasl/slx013}, \href
  {http://cdsads.u-strasbg.fr/abs/2017MNRAS.468L...6L} {468, L6}

\bibitem[\protect\citeauthoryear{Lu, Lin  \& Qin}{Lu et~al.}{2018}]{LLQ2018}
Lu W.-J.,  Lin Y.-R.,   Qin Y.-P.,  2018, \mn@doi [\mnras]
  {10.1093/mnrasl/slx176}, 473, L106

\bibitem[\protect\citeauthoryear{{McGraw} et~al.,}{{McGraw}
  et~al.}{2017}]{McGraw2017}
{McGraw} S.~M.,  et~al., 2017, \mn@doi [\mnras] {10.1093/mnras/stx1063}, \href
  {https://ui.adsabs.harvard.edu/abs/2017MNRAS.469.3163M} {469, 3163}

\bibitem[\protect\citeauthoryear{{Rupke}, {Veilleux}  \& {Sanders}}{{Rupke}
  et~al.}{2002}]{Rupke2002}
{Rupke} D.~S.,  {Veilleux} S.,   {Sanders} D.~B.,  2002, \mn@doi [\apj]
  {10.1086/339789}, \href {http://adsabs.harvard.edu/abs/2002ApJ...570..588R}
  {570, 588}

\bibitem[\protect\citeauthoryear{{Vilkoviskij} \& {Irwin}}{{Vilkoviskij} \&
  {Irwin}}{2001}]{Vilkoviskij2001}
{Vilkoviskij} E.~Y.,  {Irwin} M.~J.,  2001, \mn@doi [\mnras]
  {10.1046/j.1365-8711.2001.03985.x}, \href
  {http://cdsads.u-strasbg.fr/abs/2001MNRAS.321....4V} {321, 4}

\bibitem[\protect\citeauthoryear{{Xu}, {Arav}, {Miller}, {Kriss}  \&
  {Plesha}}{{Xu} et~al.}{2020}]{Xu2020}
{Xu} X.,  {Arav} N.,  {Miller} T.,  {Kriss} G.~A.,   {Plesha} R.,  2020,
  \mn@doi [\apjs] {10.3847/1538-4365/ab4bcb}, \href
  {https://ui.adsabs.harvard.edu/abs/2020ApJS..247...40X} {247, 40}

\bibitem[\protect\citeauthoryear{{Yao} et~al.,}{{Yao} et~al.}{2020}]{Yao2020}
{Yao} M.,  et~al., 2020, \mn@doi [\apj] {10.3847/1538-4357/ab72f3}, \href
  {https://ui.adsabs.harvard.edu/abs/2020ApJ...891...95Y} {891, 95}

\bibitem[\protect\citeauthoryear{{Yi} et~al.,}{{Yi} et~al.}{2019}]{Yi2019}
{Yi} W.,  et~al., 2019, \mn@doi [\apjl] {10.3847/2041-8213/aafc1d}, \href
  {http://adsabs.harvard.edu/abs/2019ApJ...870L..25Y} {870, L25}

\bibitem[\protect\citeauthoryear{{York} et~al.,}{{York}
  et~al.}{2000}]{York2000}
{York} D.~G.,  et~al., 2000, \mn@doi [\aj] {10.1086/301513}, \href
  {http://cdsads.u-strasbg.fr/abs/2000AJ....120.1579Y} {120, 1579}

\makeatother
\end{thebibliography}





\bsp    
\label{lastpage}
\end{document}